\documentclass[fleqn,usenatbib]{mnras}

\usepackage{newtxtext,newtxmath}

\usepackage[T1]{fontenc}

\DeclareRobustCommand{\VAN}[3]{#2}
\let\VANthebibliography\thebibliography
\def\thebibliography{\DeclareRobustCommand{\VAN}[3]{##3}\VANthebibliography}

\usepackage{graphicx}	% Including figure files

\usepackage{multicol}
\usepackage{subcaption}
\usepackage{xcolor}
\usepackage{amssymb}
\usepackage{tikz}
\usepackage{cleveref}
\usepackage{amsmath}	% Advanced maths commands
\usepackage{siunitx}
\usepackage{nth}
\usepackage{xspace}
\usepackage[normalem]{ulem}
\DeclareRobustCommand{\shortarrow}{%
\begin{tikzpicture}[baseline=-0.5ex]
    \draw[->] (0,0) -- (1ex,0);
\end{tikzpicture}
}

\crefname{figure}{Fig.}{Figs.}
\Crefname{figure}{Fig.}{Figs.}
\crefname{table}{Table}{Tables}
\Crefname{table}{Table}{Tables}
\crefname{equation}{equation}{equations}
\Crefname{equation}{Equation}{Equations}
\crefname{section}{Section}{Sections}
\Crefname{section}{Section}{Sections}
\crefname{appendix}{Appendix}{Appendices}
\Crefname{appendix}{Appendix}{Appendices}

\DeclareSIUnit \Msun {\ensuremath{\mathrm{M_\odot}}}
\DeclareSIUnit \hh {\ensuremath{\mathit{h}}}
\DeclareSIUnit \mp {\ensuremath{\mathrm{m_p}}}
\newcommand{\orcid}[1]{\href{https://orcid.org/#1}{\includegraphics[height=.7em]{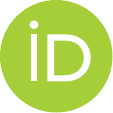}}}

\makeatletter
\DeclareRobustCommand\onedot{\futurelet\@let@token\@onedot}
\def\@onedot{\ifx\@let@token.\else.\null\fi\xspace}

\def\eg{e.g\onedot}

\def\ie{\emph{i.e}\onedot}

\makeatother
\title[Hot and cold gas accretion]{Hot gas accretion fuels star formation faster than cold accretion in high redshift galaxies}

\author[Kocjan et al]{
\orcid{0009-0003-0415-404X} Zuzanna Kocjan$^{1, 2, 3}$\thanks{E-mail: zkocjan@umd.edu},
\orcid{0000-0003-2285-0332} Corentin Cadiou$^{2}$,
\orcid{0000-0002-4287-1088} Oscar Agertz$^{2}$, and
\orcid{0000-0001-9546-3849} Andrew Pontzen$^{3}$
\\
$^{1}$Department of Astronomy, University of Maryland, College Park, MD 20742, USA\\
$^{2}$Lund Observatory, Division of Astrophysics, Department of Physics, Lund University, Box 43, SE-221 00 Lund, Sweden\\
$^{3}$Department of Physics \& Astronomy, University College London, Gower Street, London WC1E 6BT, UK\\
}

\date{Accepted XXX. Received YYY; in original form ZZZ}

\pubyear{2015}

\begin{document}
\label{firstpage}
\pagerange{\pageref{firstpage}--\pageref{lastpage}}
\maketitle

\begin{abstract}
We use high-resolution ($\simeq$ 35pc) hydrodynamical simulations of galaxy formation to investigate the relation between gas accretion and star formation in galaxies hosted by dark matter haloes of mass \SI{e12}{\Msun} at $z = 2$. At high redshift, cold-accreted gas is expected to be readily available for star formation, while gas accreted in a hot mode is expected to require a longer time to cool down before being able to form stars. Contrary to these expectations, we find that the majority of cold-accreted gas takes several hundred Myr \emph{longer} to form stars than hot-accreted gas after it reaches the inner circumgalactic medium (CGM). Approximately \SI{10}{\percent} of the cold-accreted gas flows rapidly through the inner CGM onto the galactic disc. The remaining \SI{90}{\percent} is trapped in a turbulent accretion region that extends up to $\sim 50$ per cent of the virial radius, from which it takes several hundred Myr for the gas to be transported to the star-forming disc. In contrast, most hot shock-heated gas avoids this `slow track', and accretes directly from the CGM onto the disc where stars can form.
We find that shock-heating of cold gas after accretion in the inner CGM and supernova-driven outflows contribute to, but do not fully explain, the delay in star formation.
These processes combined slow down the delivery of cold-accreted gas to the galactic disc and consequently limit the rate of star formation in Milky Way mass galaxies at $z > 2$.
\end{abstract}


%%%%%%%%%%%%%%%%%%%%%%%%%%%%%%%%%%%%%%%%%%%%%%%%%%

%%%%%%%%%%%%%%%%% BODY OF PAPER %%%%%%%%%%%%%%%%%%

\section{Introduction}\label{intro}

How gas accretes onto the central regions of the dark matter halo is of crucial importance for the formation and growth of coherent, rotating, and star-forming galactic discs. A better theoretical understanding of the processes involved, and quantification of the accretion timescales are thus key to our correct comprehension of \eg{} observations of the very high star formation rates or the formation of angular-momentum rich discs in the early Universe \citep{2023_dekel_jwst, 2023_trinca_jwst}. Recent observations from the James Webb Space Telescope (JWST) have revealed that close to \SI{60}{\percent} of galaxies have established discs at redshifts $z = 3-6$, while \SI{30}{\percent} feature disc-like morphologies as early as $z = 9$   \citep{2023_kartaltepe} with FIR emission line ([C II]-158$\mu$m) observations by ALMA revealing the existence of gas discs with high degrees of rotation as early as $z>4$ \citep[e.g.][]{Rizzo2021,RomanOliveira2023,Pope2023}.

These results are in tension with theoretical models based on contemporary cosmological simulations where galaxies are developing complex disc structures at much later times (\eg{} \citealt{2015_feng, Dubois_2016, Zavala_2016, Pillepich_2017, ferreira2022jwst}). An improved theoretical picture of how gas accretion occurs at very high redshifts is therefore crucial for our understanding of the early stages of disc galaxy formation.

\begin{figure*}
	\includegraphics[width=1.\linewidth]{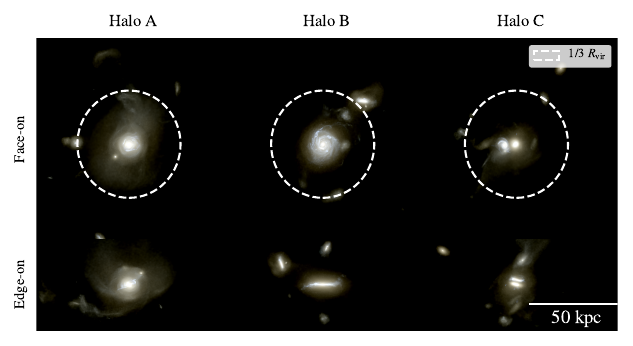}\vspace{-4mm}
    \caption{Mock images of the three galaxies in i-v-u bands mapped to r-g-b colour channels (with surface brightness between 20 and \SI{30}{mag.arcsec^{-2}}), at redshift $z \sim 2$ when $ M_{\mathrm{vir}} \approx \SI{e12}{\Msun}$. The top row shows face-on images, while the bottom row contains edge-on perspective. The dashed circle is placed at $ 1/3\, R_ {\mathrm{vir}}$. Halo A (left side) has an ellipsoidal morphology due to strong mergers happening at $z \sim 2.3$. Similarly, Halo C (right side) merges with a satellite galaxy close to $z \sim 2$. Both progenitors are disc galaxies, as seen in the edge-on image.}
    \label{fig:fig_vis}
\end{figure*}
Indeed, the details of how galaxies are supplied with gas, in particular the gas providing fuel for star formation in discs across cosmic time, are not yet fully understood. It is currently thought that at high redshift, galaxies obtain gas in a cold phase predominantly by accreting along filaments in the cosmic web \citep{2005keres,Ocvirk_2008,2009keres,2009brooks,nelson_MovingMeshCosmology_2013,cadiou_GravitationalTorquesDominate_2021}.
This filamentary accretion allows for the formation of cold ($T \sim \num{e4} - \SI{e5}{K}$) gas streams that penetrate the hot ($T \sim \num{e6}{K}$) medium in haloes with masses below \SI{e12}{\Msun}. Such streams are thought to remain cold, \ie to not experience any shock-heating. Even if shocks occur, cooling times are very short due to high gas densities \citep[\eg{}][]{2006MNRAS.368....2D}. Once the gas reaches the galactic disc, it is expected to quickly attain conditions for high star formation activity due to its low temperature \citep{Genzel_2006, Stark_2008, 2011_SF_rodighiero, chen_SF_2021} and high density, directly contributing to the growth of the disc (\eg{} \citealt{2009Natur.457..451D, 2009MNRAS.397L..64A, 2012ApJ...745...11G,2020MNRAS.497.4346K}). Some work using cosmological simulations has shown that once the streams enter the region below $ 1/3\, R_ {\mathrm{vir}}$, they typically join a region of a complex structure, in which the gas mixes and resides before arriving in the central galactic disc, usually inside $\sim 1/10\,R_{\mathrm{vir}}$ \citep{2010_ceverino}. Similarly, other simulations predict the existence of `cold flow discs', which orbit the central galaxy as an extension of the galactic disc (\eg{} \citealt{stewart_2011, stewart_2013, Danovich_2015, 2019_dekel}), before delivering fuel for star formation inside the disc.

If the mass of the halo is above \SI{e12}{\Msun}, at low redshifts ($z \lesssim 2$), the accreted gas is likely to experience virial shock-heating on its way to the galaxy centre. The gas temperature will rise to the virial temperature $T_{\mathrm{vir}}$ of the halo at a distance close to $R_{\mathrm{vir}}$ away from the galaxy \citep{accretion_modes_vandeVoort}. For a galaxy with a mass similar to that of the Milky Way, this is close to \SI{e6}{K} \citep{2005keres}. Subsequently, the hot-accreted gas will enter the CGM and arrive at the galactic disc. Some models suggest a connection between the hot accretion mode and the formation of thin discs at low redshifts in more massive haloes \citep{Sales_2012, Stern_2021, 2023_sijie,2023_stern}. \citet{2022_hot_acc_hafen} find that for thin disc formation to be possible, a virialised CGM has to form, in which `hot flows' deliver gas to the disc. This hypothesis is qualitatively in agreement with observations of thin disc morphologies, as they are more prevalent in high-mass, star-forming galaxies at low-z \citep{Kassin_2012, simons_2017}. However, some models \citep[\eg{}][]{Muratov_2015} show that the formation of coherent, star-forming discs can be also linked with dissolution of strong galactic outflows, rather than being a byproduct of the hot accretion mode. 

How quickly gas becomes available for star formation after entering the galactic halo will ultimately depend on the mode of accretion. Currently, theoretical models assume that cold accretion is the main source of star-forming gas at high redshifts \citep{Genzel_2006, Stark_2008, 2011_SF_rodighiero, chen_SF_2021}. The cold flows are expected to rapidly provide gas that is readily available for star formation, as opposed to hot-accreted gas which requires more time to cool down and condense. However, as we demonstrate in this work, how cold and hot gas flows through dark matter haloes into star-forming discs is more complex.

In this paper, the goal is to understand the timescales involved for halo gas to accrete onto galaxies and form stars, and how this depends on the various modes of gas accretion. To that end, we make use of a set of zoom simulations carried out with the adaptive mesh refinement (AMR) code RAMSES \citep{RAMSES}. Thanks to the use of tracer particles \citep{cadiou_AccurateTracerParticles_2019}, we can access the Lagrangian trajectory of gas. This setup allows us to study precisely the thermal and kinematic evolution of gas in each of the accretion modes (hot and cold) and measure how long it takes for the accreted gas to form new stars in the main galaxy.

The paper is structured as follows. \cref{metho} details the method, including the numerical setup and classification criteria for gas subsets. In \cref{results}, we present the results of our numerical analysis and compare these to the theoretical predictions. The discussion and conclusions are provided in \cref{discus}.

\section{Methods} \label{metho}

\subsection{Numerical setup} \label{setup}

\begin{figure*}
	\includegraphics[width=2.05\columnwidth]{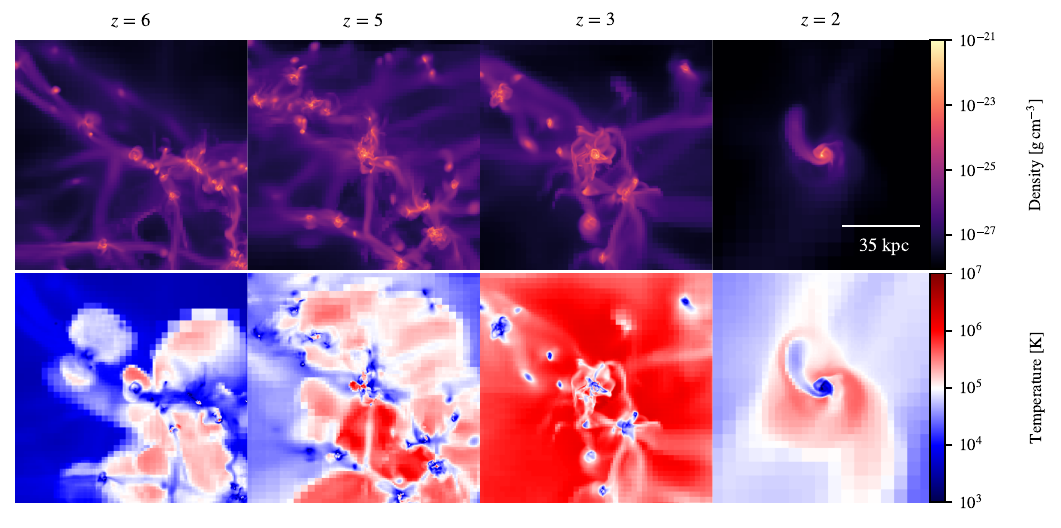}\vspace{-4mm}
    \caption{Projections from the simulations showing gas density (top row) and temperature (bottom row) of galaxy A at different redshifts (as labelled). The box side length is \SI{100}{kpc}, the depth of the projection is \SI{300}{kpc}. Cold-accreted gas is delivered to the galaxy via filamentary accretion, which dominates at higher redshifts. The hot phase enters the halo spherically and is more prevalent at lower \textit{z}. At $z \geq 5$, an accretion region can be seen forming around the central galaxy. At $z = 2$, that region is composed mostly of cold-accreted gas, while a hot halo is feeding the galaxy with gas from larger distances.}
\label{fig:projections}
\end{figure*}
The analysis was conducted using three hydrodynamical simulations performed with the code RAMSES \citep{RAMSES} with Monte-Carlo tracer particles \citep{cadiou_AccurateTracerParticles_2019}. The spatial resolution of these simulations is \SI{35}{pc} and is reached in star forming regions, with a comoving box size of length \SI{100}{Mpc.\hh^{-1}}. The dark matter particle mass is \SI{1.6e6}{\Msun} and the stellar mass \SI{1.1e4}{\Msun}. More technical details on this numerical setup can be found in \citet{CC_stellar_angular}.

Briefly, the simulations use the subgrid physics of New Horizon \citep{dubois_2021_NH}. This model has been shown to reproduce the stellar mass-to-halo mass relation at low redshift \citep{dubois_2021_NH}. The implemented star formation (SF) recipe consists of: \textit{(i)} SF allowed above a density of $n_{\mathrm{0}} = \SI{10}{cm^{-3}}$ following a SF recipe based on supersonic turbulence simulations \citep{Kimm_2017, 2017_trebitsch, Trebitsch_2021}; \textit{(ii)} stellar populations sampled with a \citet{kroupa2001} initial mass function; and \textit{(iii)} a mass loss fraction from supernovae explosions $\eta_{\mathrm{SN}}$ equal to \SI{32}{\percent} (with a metal yield of 0.05). Modelling of Type II supernovae follows the mechanical feedback model presented in \citet{Kimm_2017}, with an additional boost in momentum to approximate the effect of pre-supernova feedback\footnote{\citet{Geen_2015} demonstrated that the terminal momentum from SNe is increased by including photoionization from massive stars. The ionization process overpressurizes and decreases the density of the surroundings into which SNe explode, leading to a stronger coupling and more momentum generation by the hot SN bubble. As we do not include radiative transfer in our models, nor would we be able to resolve most Str\"omgren spheres at the adopted spatial resolution, we include the effect of photoionization in the terminal momentum of SNe following \citet{Kimm_2017}.} \citep{Geen_2015}. AGN feedback follows the model of \citet{dubois_2012}, where the Eddington--limited Bondi--Hoyle--Lyttleton accretion rate is tracked --- at high accretion rates, the feedback model injects energy into the AMR grid (`quasar mode'), while at smaller accretion rates, feedback is injected as momentum in a bipolar jet (`radio mode').

To assess the importance of feedback processes, we include in \cref{NUT} an analysis of outputs from the `NUT' zoom simulation \citep{powell_ImpactSupernovadrivenWinds_2011}, which omits both AGNs and supernovae feedback \citep{Rams_y_2021}. That section includes both technical details on this setup, and key results obtained using this simulation.

We analyse simulations of three galaxies and their evolution between redshifts $ 2 < z < 15$. At redshift $z=3$ and $2$, the halo mass of the galaxies are \SI{3e11}{\Msun} and \SI{e12}{\Msun} respectively, while the tracer particle mass is $\SI{2.3e5}{\Msun}$. In \cref{fig:fig_vis}, we show the three galaxies face-on and edge-on: Halo A, B, and C respectively, at $z \sim 2$. Halo A (left side) merges with a galaxy around $z \sim 2.3$, hence the more spherical morphology. Halo C (right side) is shown while undergoing a merger with a satellite galaxy. Despite showing different morphologies at $z \sim 2$, as seen in \cref{fig:fig_vis}, the galaxies show the same behaviour in terms of accretion and star formation throughout the entire duration of the simulation, thus making our results invariant of the structural properties of galaxies. In \cref{fig:projections}, we present density (top row) and temperature (bottom row) density--weighted projections of the gas surrounding halo A at redshifts $z \sim 2 - 6$, with several satellite galaxies visible in the cosmic web. Note that the simulated galaxies are more compact than observed ones by a factor of approximately two \citep[as reported in \eg{} fig. 11 of ][]{dubois_2021_NH}.

\subsection{Cold and hot accretion definitions} \label{ext}
In the following, we refer to the virial radius of the dark matter halo as $R_{\mathrm{vir}}$, the virial mass as $M_{\mathrm{vir}}$, and the virial temperature as $T_{\mathrm{vir}}$. We analyse all tracer particles that have crossed  $2\, R_ {\mathrm{vir}}$ and entered the inner CGM, defined as the region below $ 1/3\, R_ {\mathrm{vir}}$ from the galactic centre. 

The galactic disc is defined as the region below 1/10 $R_{\mathrm{vir}}$. For each time output we use the respective $R_ {\mathrm{vir}}$ value. Therefore, $R_ {\mathrm{vir}}$ corresponds to a different distance when the accreted particle crosses $2\, R_ {\mathrm{vir}}$, $ 1/3\, R_ {\mathrm{vir}}$ and 1/10 $R_{\mathrm{vir}}$, since it grows with time. The `cold gas' accreted via the dense filaments refers to baryons that:

\begin{enumerate}
    \item are accreted, \ie reside in the inner CGM, within $r < 1/3 \,R_{\mathrm{vir}}$ of the halo at the end of the simulation ($z= 2$);
    \item are in the star phase at the end of the simulation;
    \item are in the gas phase before crossing $ 1/3\, R_ {\mathrm{vir}}$ (which excludes stars formed ex-situ);
    \item do not have a density higher than \SI{1}{\mp.cm^{-3}} after crossing $2\, R_ {\mathrm{vir}}$ and before entering the inner CGM. This allows us to exclude dense gas transported to the galaxy via satellite galaxies\footnote{This value is above the range of the density profile of cold gas streams, as well as below the star formation recipe threshold in these simulations.};
    \item are not heated above the temperature threshold $ T_{\mathrm{cold}} = {\SI{3e4}{K}}$ after crossing  $2\, R_ {\mathrm{vir}}$ and before entering the inner CGM.
\end{enumerate}

Similarly, we classify inflowing gas as `hot' if it meets criteria (\textit{i}) -- (\textit{iv}), and reaches a temperature above $ T_{\mathrm{hot}} = \SI{5e5}{K}$ at any point after crossing $2\, R_ {\mathrm{vir}}$ and before crossing $ 1/3\, R_ {\mathrm{vir}}$ for the first time.

In criterion \textit{(v)}, we use the cold gas threshold $T_{\mathrm{cold}}$ to select all the filamentary-accreted gas as it is the maximum temperature the filaments are expected to reach inside the galactic halo due to photo-heating from UV background \citep{Goerdt_2010, Mandelker_2020}, assuming gas self-shielding above \SI{e-2}{cm^{-3}}. For hot gas classification, we set $ T_{\mathrm{hot}}$ as the minimum temperature achieved by quasi-spherical accretion \citep{Mandelker_2020} as it corresponds to one-half of the virial temperature $T_{\mathrm{vir}}$ of a Milky Way-like galaxy\footnote{For a galaxy with virial mass $M_{\mathrm{vir}} \sim \SI{e12}{\Msun}$, $ T_{\mathrm{vir}} \sim \SI{e6}{K}$ \citep{2009Natur.457..451D}.}, where $T_{\mathrm{vir}}$ is assumed to be the typical temperature of the shock-heated gas.

We note that the above classification of cold and hot-accreted gas excludes gas that is warmer than $T_{\mathrm{cold}}$ but cooler than $T_{\mathrm{hot}}$. The physics governing the intermediate category is likely complex and may include gas which is disrupted within a cold flow through hydrodynamical instabilities \citep[see \eg{}][]{mandelker_2016,padnosInstabilitySupersonicCold2018,aung_KelvinHelmholtzInstabilitySelfgravitating_2019,Mandelker_2020}.

\subsection{Accretion timescales}

Using the Lagrangian history, for each gas tracer we define $t_{2}$ and $t_{1/3}$ as the time at which gas crosses respectively $ 2\, R_ {\mathrm{vir}}$ and $ 1/3\, R_ {\mathrm{vir}}$ for the first time, and $t_{\star}$ as the time when gas forms stars. This allows us to define two timescales involved in the process of gas accretion and star formation: the \textit{accretion} timescale $t_{2\shortarrow 1/3}$ and the \textit{conversion} timescale $t_{1/3\shortarrow\star}$. The first one, $t_{2\shortarrow1/3}$, is defined as the difference between $t_{2}$ and $t_{1/3}$. This timescale quantifies the travelling time of gas through the galactic halo to the inner CGM. Similarly, the conversion timescale, $t_{1/3\shortarrow\star}$, is defined as the difference between $t_{1/3}$ and $t_{\star}$. This quantity determines how quickly gas forms stars after it enters the inner CGM.

\section{Results} \label{results}

\begin{figure}
	\includegraphics[width=0.95\columnwidth]{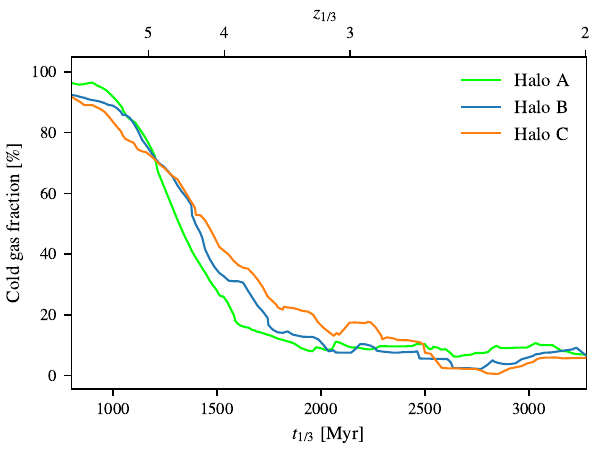}
    \caption{Evolution of cold accretion rate as a percentage of the total accretion for all three galaxies studied. The percentage decreases in time --- at redshift $\sim 5$, around \SI{90}{\percent} of the total accretion consists of cold gas inflow, while at redshift $z \sim 2$, when ${M_{\mathrm{vir}} = \SI{e12}{\Msun}}$, it is close to \SI{10}{\percent}. The percentage is counted at $t_{1/3}$, the time when gas crosses $ 1/3\, R_ {\mathrm{vir}}$.}
    \label{fig:acc}
\end{figure}

As discussed in the Introduction, a range of studies have previously shown that filamentary accretion dominates at higher redshifts, while hot accretion becomes the main source of gas at lower $z$. To study when this transition takes place and to measure the accretion rate for cold-accreted gas, we select all of the gas which falls below $ 1/3\, R_ {\mathrm{vir}}$ until the end of the simulation at $z = 2$ (without necessarily forming stars) and compute the percentage of cold-accreted gas to the total accreted gas as a function of $t_{1/3}$. The results are shown in \cref{fig:acc}. We recover the expected behaviour that, until $t_{1/3} = \SI{1.2}{Gyr}$ ($z_{1/3} \sim 5$), around \SI{90}{\percent} of the gas below $ 1/3\, R_ {\mathrm{vir}}$ originates from cold accretion. The rate drops to around \SI{10}{\percent} close to $t_{1/3} = 3.2\,\mathrm{Gyr}$ $(z_{1/3} \sim 2)$. These results are in broad agreement with \citet{2005keres, 2006MNRAS.368....2D, Ocvirk_2008, 2009keres}, who found a transition from cold to hot-dominated accretion around a halo mass of $\sim \num{5e11} -  \SI{e12}{\Msun}$ at $z \sim 2 - 3$. In our simulation, $M_{\mathrm{vir}} = \SI{3e11}{\Msun}$ at $z \sim 3$, and ${M_{\mathrm{vir}} = \SI{e12}{\Msun}}$ at $z \sim 2$.

\begin{figure}
	\includegraphics[width=0.95\columnwidth]{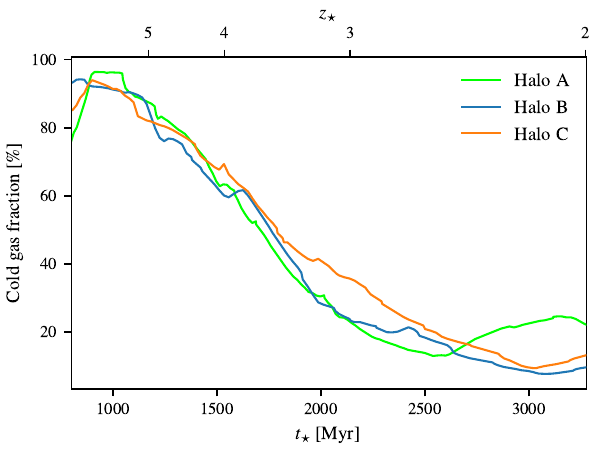}
    \caption{Evolution of the star formation rate of cold-accreted gas in all three simulations as a percentage of the total star formation rate. The percentage decreases in time as hot accretion becomes more dominant at lower $z$.}
    \label{fig:fig_sf_rate}
\end{figure}

\cref{fig:fig_sf_rate} shows the corresponding results measured at star formation time $t_{\star}$ (as opposed to $t_{1/3}$, as in \cref{fig:acc}) for each gas population in all three simulations. The percentage decreases in time --- at redshift $z_{\star} \sim 5$, around \SI{90}{\percent} of the total star formation is regulated by cold gas inflow, while at a redshift $z_{\star} \sim 2$, SFR is fuelled in around \SI{20}{\percent} by cold filaments falling onto the galactic disc. The percentage was counted at $t_{\star}$, when gas forms stars below $ 1/3\, R_ {\mathrm{vir}}$. We consider only tracer particles which are in the form of stars at the end of the simulation.

\begin{figure}
	\includegraphics[width=0.95\columnwidth]{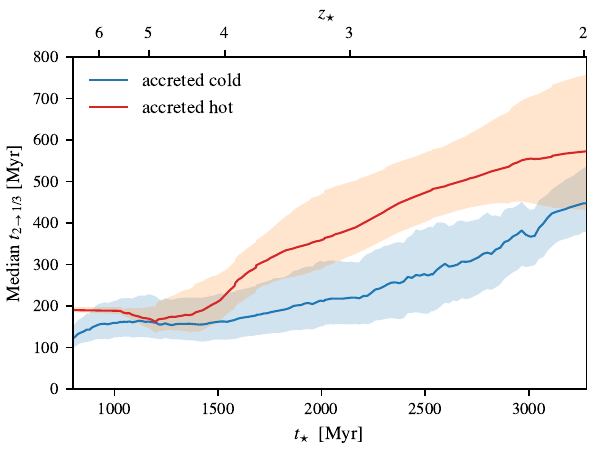}
    \caption{Median accretion timescale $t_{2\shortarrow1/3}$ (solid lines) for cold- and hot-accreted gas versus their star formation time $t_{\star}$ in halo A. Shaded regions encompass the \nth{75} and \nth{25} percentiles.} Gas accreted via cold filaments reaches the inner CGM faster than hot gas accreted quasi-spherically --- the difference between the medians reaches up to \SI{200}{Myr}.
    \label{fig:timescales_2_13}
\end{figure}

The cold streams are expected to penetrate the hot galactic halo and deliver gas into the inner CGM in less time than the hot accretion channel \citep{dubois_2012, 2015_goerdt}. We confirm this in \cref{fig:timescales_2_13} by showing the median accretion timescale $t_{2\shortarrow1/3}$ of hot- and cold-accreted gas versus their star formation time  $t_{\star}$, in halo A. The timescales become longer at lower redshifts as the galactic halo grows in mass, thereby increasing $R_{\mathrm{vir}}$. Additionally, the free-fall time of the galactic halo over-density increases as the average density of the Universe decreases with time. Similar results are obtained for the other two galaxies.

\begin{figure*}
    \includegraphics[width=\linewidth]{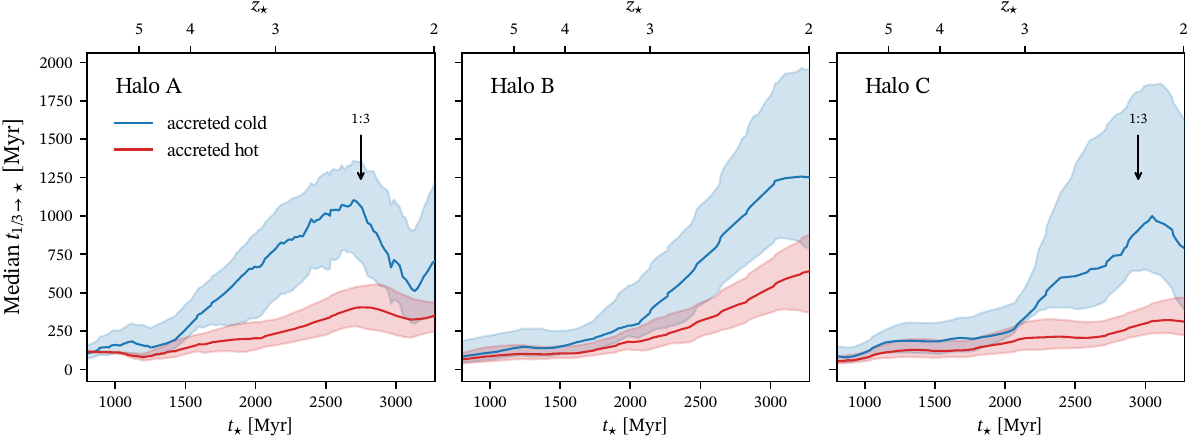}
    \caption{Median conversion timescale $t_{1/3\shortarrow\star}$ for cold and hot-accreted gas versus their star formation time $t_{\star}$. Cold-accreted gas experiences a delay and takes more time to form stars than hot-accreted gas, with the difference being significantly larger at redshifts below 4. The drop close to redshift $\sim 2.3$ in halo A (left) and the multiple declines in halo C (right) coincide with mergers (the mass ratios of the merging galaxies above redshift $z_{\star} \sim 3$ are shown), resulting in starbursts and thus reducing $t_{1/3\shortarrow\star}$. Halo B only undergoes minor mergers, that do not affect $t_{1/3\shortarrow\star}$ significantly.}
    \label{fig:fig_combined}
\end{figure*}

However, the above results \emph{do not} mean that cold-accreted gas can reach the central galaxy and form stars faster than hot-accreted gas. \cref{fig:fig_combined} shows the evolution of the median conversion timescale $t_{1/3\shortarrow\star}$ for all three galaxies. In all systems, the median time for cold accretion is always longer than for hot accretion throughout the entire duration of the simulation. For cold-accreted gas, $t_{1/3\shortarrow\star}$ increases greatly at redshifts $z_{\star}$ below $\sim 4$, becoming several times higher than the conversion timescale for hot-accreted gas. In halo A, at $z_{\star} \sim 2.5$, the difference between the two medians reaches around \SI{800}{Myr}, while at $z\gtrsim 5 $ the delay is not as significant (up to \SI{100}{Myr}), which we will later show to be connected to the formation of an accretion region (visible in \cref{fig:projections}). Similarly, in halo B, the medians grow over time and cold-accreted gas reaches $t_{1/3\shortarrow\star} \sim \SI{1.25}{Gyr}$ at $z_{\star} \sim 2$, while hot-accreted gas arises up to $t_{1/3\shortarrow\star} \sim \SI{600}{Myr}$.

\begin{figure}
	\includegraphics[width=\columnwidth]{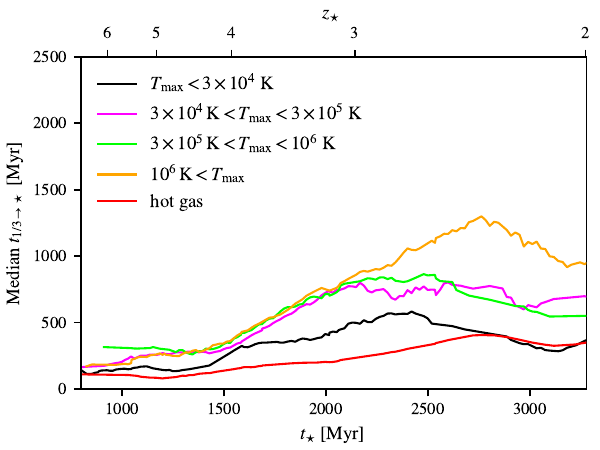}
    \caption{Median $t_{1/3\shortarrow\star}$, same as in \cref{fig:fig_combined}, here for cold-accreted gas in halo A binned into 4 subsets, based on the maximum temperature reached by the tracer particles after crossing $ 1/3\, R_ {\mathrm{vir}}$. The higher the temperature that gas reaches after entering the inner CGM, the longer the conversion timescale $t_{1/3\shortarrow\star}$ to form stars is. Possible shock-heating below $ 1/3\, R_ {\mathrm{vir}}$ could therefore be one of the causes of the delay. However, the median line for gas which stays cold is still above the hot-accreted gas line, hence suggesting an additional slowing factor.}
    \label{fig:fig_groups_temp}
\end{figure}

In \cref{fig:fig_combined}, there is a sharp drop in the median curve for cold-accreted gas in halo A at $z\sim2.3$, which coincides with a major merger. A similar evolution takes place in halo C, whereas halo B undergoes minor merging with nearby galaxies between $z\sim 2.1$ and $\sim 2$. We interpret the drops as merger-induced bursts that deplete the gas reservoir; observations (\eg{} \citealt{2012_scudder, 2015_knapen}), as well as simulations (\eg{} \citealt{2010_teyssier, 2013_moreno, 2022_segovia_otero, 2022_renaud}), have shown that galaxy interactions lead to gas compression and increased star formation activity. We note that despite the dense gas from satellite galaxies being excluded from our analysis (by implementing a density criterion), some of the less dense, cold-accreted gas orbiting close to the satellites may still effectively accelerate star formation in the central galaxy after merging. A more detailed analysis will be required to confirm this, which is beyond the scope of this work.

We check the robustness of these results by modifying the temperature thresholds implemented for cold- and hot-accreted gas classification in ranges $\SI{e4}{K} < T_{\mathrm{cold}} < \SI{e5}{K}$ and $\SI{e5}{K} < T_{\mathrm{hot}} < \SI{e6}{K}$. Decreasing $T_{\mathrm{hot}}$ and increasing $T_{\mathrm{cold}}$ reduces the difference between the two median curves at redshifts $z_{\star} < 3$ only to a small extent --- a decrease of at most \SI{100}{Myr}. In all cases, we obtain the same conclusion: gas transported by cold streams into the CGM takes longer to form stars, compared to gas accreted via the hot accretion channel. We find similar behaviour in the NUT galaxy; the results are described in more detail in \cref{NUT}.

\subsection{Origin of the star formation delay}

We next investigate four mechanisms potentially contributing to the delay in star formation experienced by cold-accreted gas: shock heating, feedback outflows, slow angular momentum dissipation, and geometry of accretion and discuss the role of each in turn. Since we are investigating what affects $t_{1/3\shortarrow\star}$, we analyse heating and outflows of gas that take place after gas arrives inside the inner CGM ($r < {1/3 \,R_{\mathrm{vir}}}$). Additionally, we study how the angular momentum dissipation and stability of the cold flows affect the kinematic evolution in the $\sim \SI{600}{Myr}$ before star formation, which may include gas at ${r > 1/3\,R_{\mathrm{vir}}}$.

To be able to track the relation between these properties and the conversion timescales, we make an additional classification of cold-accreted gas into `fast' and `slow', depending on whether the conversion timescale $t_{1/3\shortarrow\star}$ is lower or higher than the median $t_{1/3\shortarrow\star}$ for hot-accreted gas, at a specific star formation time $t_{\star}$ (\cref{fig:fig_combined}). We note that fast cold-accreted gas only contains \num{10} -- \SI{15}{\percent} of the total amount of cold-accreted gas accreted throughout the entire simulation. % --- the vast majority of it has higher $t_{1/3\shortarrow\star}$ than the median hot-accreted gas $t_{1/3\shortarrow\star}$ at a given $t_{\star}$.

\paragraph*{Shock-heating.}
With the adopted definitions (\cref{ext}), cold-accreted gas may still undergo shock-heating below  $1/3\, R_ {\mathrm{vir}}$, as the adopted temperature criterion only concerns gas between $2\, R_ {\mathrm{vir}}$ and $1/3\, R_ {\mathrm{vir}}$. Shock-heating close to the disc would suppress star formation due to longer cooling times. This gas heating can be caused by \eg{} shock-heating of inflowing matter, supernovae explosions, or gas compression inside the dense inner CGM.

To understand if heating below  $1/3\, R_ {\mathrm{vir}}$ can delay star formation, we sort cold-accreted gas into four subgroups, depending on the maximum temperature{, $T_\mathrm{max}$}, each particle reaches after having crossed $ 1/3\, R_ {\mathrm{vir}}$ (before star formation). In \cref{fig:fig_groups_temp}, we present the same measurements as in \cref{fig:fig_combined} for halo A, but for the different subgroups. The temperature ranges are as follows: $T_{\mathrm{max}} < T_{\mathrm{cold}} = \SI{3e4}{K}$ (black line), $\SI{3e4}{K} < T_{\mathrm{max}} < \SI{3e5}{K}$ (magenta line), $\SI{3e5}{K} < T_{\mathrm{max}} < \SI{e6}{K}$ (lime line) and $T_{\mathrm{max}} >  \SI{e6}{K}$ (orange line). Approximately \SI{40}{\percent} of the cold-accreted gas experiences heating above $T_{\mathrm{cold}}$ in the inner CGM. As seen in the figure, the higher the temperature reached, the longer it takes for the gas to form stars. However, for cold-accreted gas that does remain cold ($T<T_{\mathrm{cold}}$, \SI{60}{\percent} of the cold-accreted gas population), the median timescale $t_{1/3\shortarrow\star}$ is still higher than the median $t_{1/3\shortarrow\star}$ for hot-accreted gas for the majority of $t_{\star}$.
Therefore, while part of the star formation delay experienced by cold-accreted gas is due to heating after accretion in the inner CGM, another process is required to fully explain the delay for the gas that remains cold all the way to the star forming disc.

\begin{figure}
	\includegraphics[width=\columnwidth]{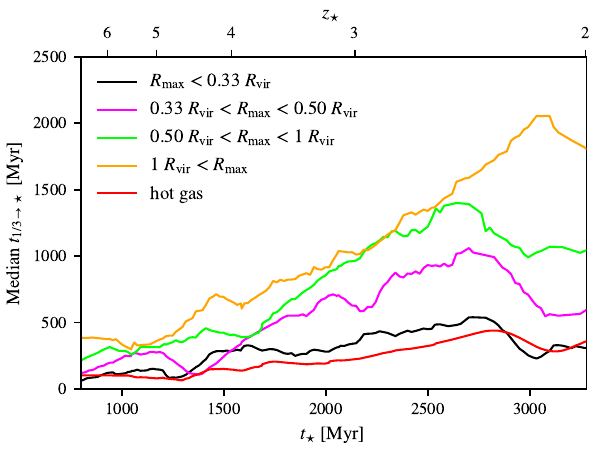}
    \caption{Same as \cref{fig:fig_groups_temp}, except the sorting criterion is changed to the maximum distance (in $R_{\mathrm{vir}}$) reached by particles after having crossed $ 1/3\, R_ {\mathrm{vir}}$ for the first time. The further the gas is ejected, the longer it takes to return to the inner CGM and to form stars. However, the median $t_{1/3\shortarrow\star}$ for cold-accreted gas that does not leave the circumgalactic medium still lies above the median for hot-accreted gas, suggesting an additional contributor to the delay.}
    \label{fig:fig_groups_dist}
\end{figure}

\paragraph*{Outflows.}
To investigate the effect of galactic outflows on $t_{1/3\shortarrow\star}$, we retain the cold-accreted gas definition and conduct a similar analysis as above, focusing not on temperature but on the maximum distance{, $R_\mathrm{max}$,} reached by gas after crossing $ 1/3\, R_ {\mathrm{vir}}$ for the first time. \cref{fig:fig_groups_dist} shows the median $t_{1/3\shortarrow\star}$ for the following distance ranges: $R_{\mathrm{max}} < 1/3\,R_{\mathrm{vir}}$ (black line, \SI{41}{\percent} of the cold gas population), $1/3\,R_{\mathrm{vir}} < R_{\mathrm{max}} < 1/2\,R_{\mathrm{vir}}$ (magenta line, \SI{39}{\percent}), $1/2\,R_{\mathrm{vir}} < R_{\mathrm{max}} < R_{\mathrm{vir}}$ (lime line, \SI{12.5}{\percent}) and $R_{\mathrm{max}} > R_{\mathrm{vir}}$ (orange line, \SI{7.5}{\percent}). Only \SI{41}{\percent} of cold gas accreted throughout the entire simulation duration remains inside the inner CGM ($<1/3\,R_{\mathrm{vir}}$) before star formation \footnote{In the NUT simulation (\cref{NUT}), \SI{85}{\percent} of cold-accreted gas remains inside the inner CGM, due to feedback processes not being included in the simulation.}. When it comes to hot gas, around 80\% of it remains inside the CGM after accretion and before star formation, while 97\% of it stays inside $0.5\, R_ {\mathrm{vir}}$.

As we show in \cref{fig:fig_groups_dist}, the further the gas is expelled, the longer the delay. However, even the cold-accreted gas that never flows out of the inner CGM after being accreted experiences a delay compared to the hot gas, albeit a much smaller one.
The same conclusions are obtained using the other two simulations with New Horizon physics, as well as the NUT simulation with no feedback included. We note that the vast majority of cold accreted gas ($\sim 80\%$) is accounted for by the black and magenta lines, \ie gas that can travel out to $0.5\, R_ {\mathrm{vir}}$ before ending up in the star forming galaxy.  As we demonstrate below, this is is not necessarily due to cold gas being entrained in galactic outflows, but rather a signature of its complex accretion geometry. 

\begin{figure}
    \includegraphics[width=\columnwidth]{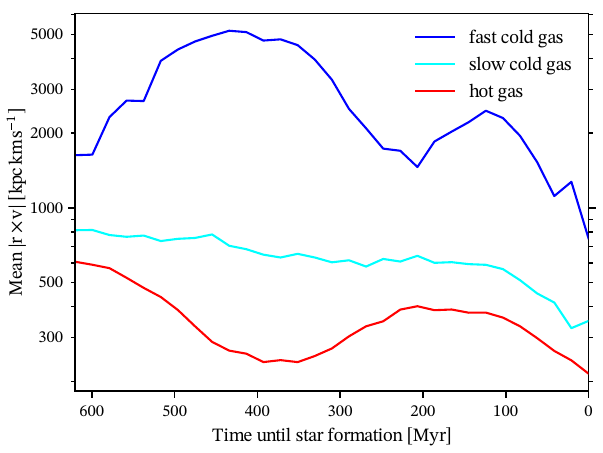}
    \caption{Mean specific angular momentum $j_\mathrm{gas}$ of fast and slow cold-accreted gas, together with hot-accreted gas, up to \SI{600}{Myr} before star formation. High angular momentum is unlikely to be the cause of the delay in star formation of cold-accreted gas, as slow gas resides in an environment of low angular momentum before star formation, while fast cold-accreted gas arrives at the disc with a much higher amount of $j$ and loses the majority before star formation (around \SI{100}{Myr}). Hot-accreted gas does not carry as much angular momentum as filamentary-accreted gas.}
    \label{fig:fig_AM}
\end{figure}

\begin{figure*}
\begin{multicols}{2}
    \includegraphics[width=0.92\linewidth]{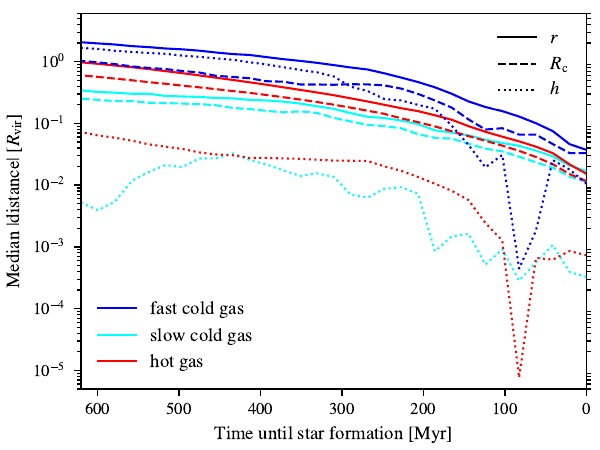}\par
    \includegraphics[width=0.92\linewidth]{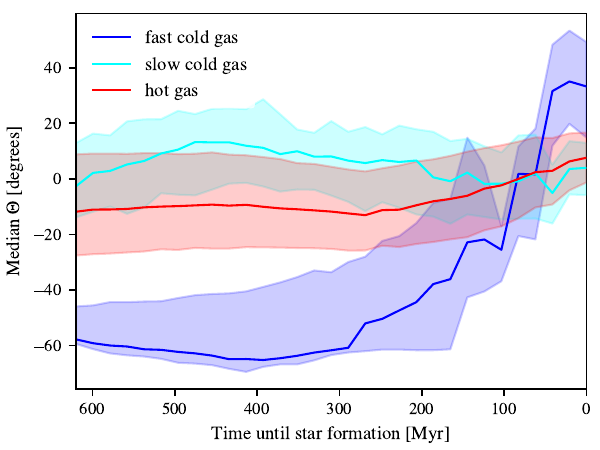}\par
\end{multicols}\vspace{-5mm}
\caption{Evolution of the median of the angle $\Theta$ (right panel) and the absolute values of distance $r$, height $h$ and cylindrical radius $R_{\mathrm{c}}$ (solid, dotted, and dashed lines respectively, on the left panel) before star formation in halo A. The shaded region shows the difference between the 40th and 60th percentile. All of the variables presented show that the majority of cold gas accreted via the filaments resides in an area around $0.25\, R_ {\mathrm{vir}}$ away from the galaxy for a few hundred Myr before entering the inner-dense disc region below $ 1/10\, R_ {\mathrm{vir}}$, and later the active star formation area. The angle between accretion and the galactic plane (right side), shows that the structure containing slow cold-accreted gas is lingering at a low angle relative to the galactic plane. A fraction of cold-accreted gas that manages to turn into stars quickly accesses the area from above or below the dense region and undisturbed quickly forms stars. The accreted quasi-spherically hot gas arrives at the disc from all directions, only mixing with cold-accreted gas in the dense extended disc-like structure below $ 1/10\, R_ {\mathrm{vir}}$. The median angle for hot-accreted gas decreases from \ang{10} to around \ang{0} while approaching star formation, which agrees with an isotropic model of hot accretion. As opposed to slow cold-accreted gas, hot-accreted gas does not linger in the region at around $0.25\, R_ {\mathrm{vir}}$.}
\label{fig:geometry}
\end{figure*}

\paragraph*{Angular momentum.}

We have shown that both heating and outflows may delay star formation, although neither can fully explain the delay on their own.
To test the hypothesis that the delay is due to cold-accreted gas entering the inner CGM with higher angular momentum, we compute the mean of absolute values of the specific angular momentum of the fast and slow cold gas, as well as the hot-accreted gas. We define the mean specific angular momentum as $j_{\mathrm{gas}} =  \langle|\textbf{r} \times \textbf{v}|\rangle$, where the mean runs over all selected particles 
and where $\textbf{v}$ is the velocity of each particle and $\textbf{r}$ is the position vector, both with respect to the galaxy. We analyse gas that forms stars at redshifts $2 < z < 3$, where the difference in $t_{1/3\shortarrow\star}$ {between cold- and hot-accreted gas} is the most significant (\cref{fig:fig_combined}). We refer to \cite{cadiou_GravitationalTorquesDominate_2021} for a more in-depth analysis of the distribution of the angular momentum of the accreted gas for galaxies of similar mass with the same model.

We show in \cref{fig:fig_AM} the $j_{\mathrm{gas}}$ versus time up to \SI{600}{Myr} before star formation. Throughout the \SI{600}{Myr}, the specific angular momentum of the slow gas remains below \SI{e3}{kpc.km.s^{-1}}; starting from $j_{\mathrm{gas}} \sim  \SI{8e2}{kpc.km.s^{-1}}$, it slowly loses angular momentum before entering the star-formation region.
Comparatively, $j_{\mathrm{gas}}$ of the fast cold gas is several times larger. The mean for both slow and fast cold gas is always higher than the mean $j_{\mathrm{gas}}$ of hot-accreted gas. These results show that slow gas is not trapped in an angular-momentum-rich region; on the contrary, it is the fast-accreting gas that brings a large amount of angular momentum to the star-forming region. This leads to the seemingly peculiar result that gas that arrives rapidly to the star forming disc can either be rich in a specific angular momentum ('fast cold gas') or poor ('hot gas'), whereas a slow down occurs for cold gas with an intermediate specific angular momentum content. This can be explained by the accretion geometry, which we turn to next.

\begin{figure*}
	\includegraphics[width=1.9\columnwidth]{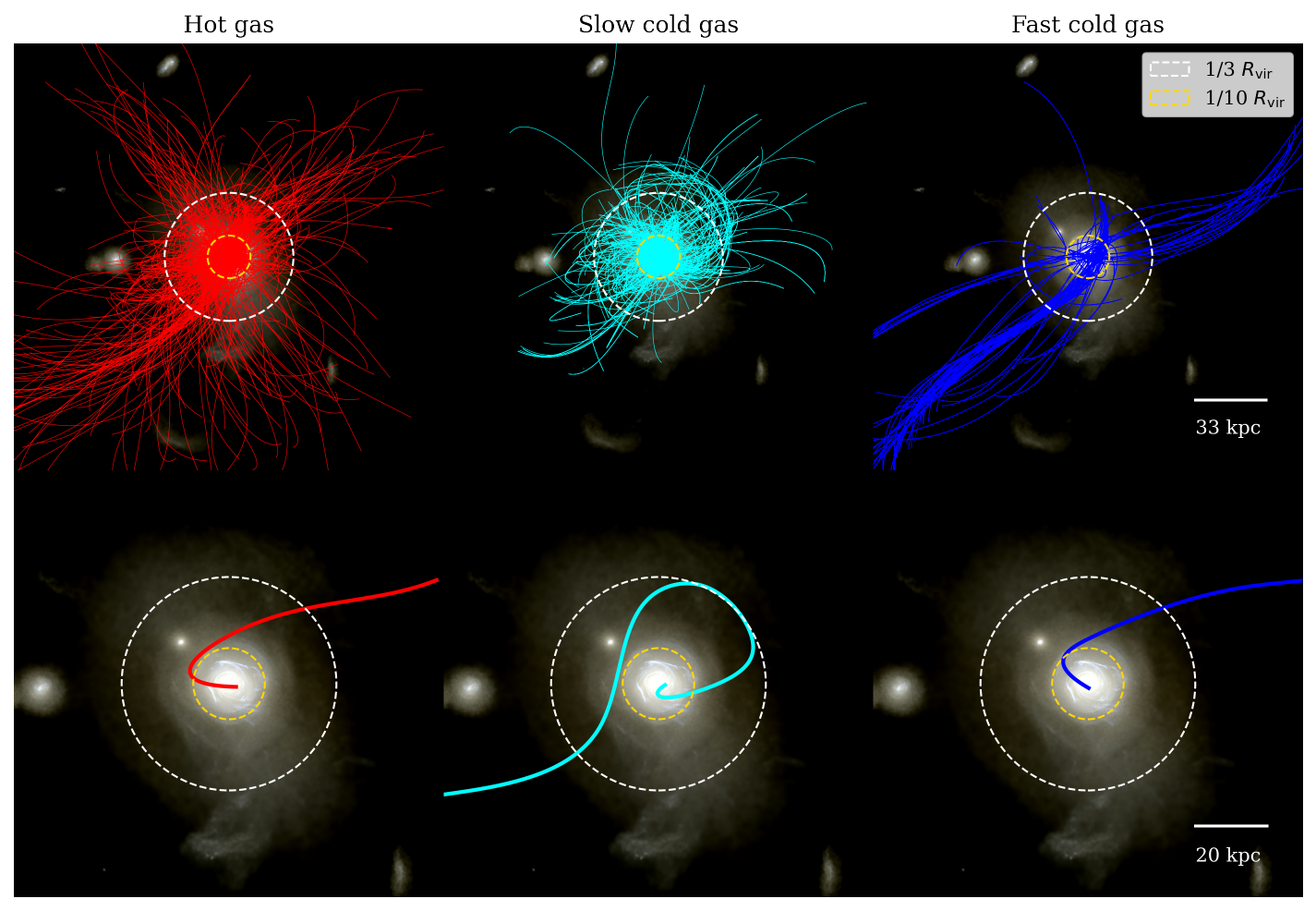}\vspace{-2mm}
    \caption{Trajectories of 500 tracer particles selected randomly from each subset, on top of {mock images of stellar emission in the u, v and b bands of} halo A (top row). Hot-accreted gas (top-left) enters the inner CGM from various directions and quickly joins the disc formed below 1/10  $R_{\mathrm{vir}}$, where it resides before forming stars. Slow-cold gas (top-middle) resides in the inner CGM in an extended spiral/disc/cloud structure before joining the galactic disc for more than a few hundred Myr. Fast-cold gas (top-right) travels through the structure and arrives directly in the galactic disc, quickly providing fuel for star formation. The bottom row shows the trajectory of one randomly selected particle from each subgroup. All the tracer particles presented are with $t_{\mathrm{\star}}$ between $2 < z < 3$. Dashed circles show the radii $ 1/3\, R_ {\mathrm{vir}}$ and $ 1/10\, R_ {\mathrm{vir}}$. The trajectories show tracer positions up to \SI{400}{Myr} before star formation. }
    \label{fig:fig_traj}
\end{figure*}

\paragraph*{Accretion geometry.}
While cold gas fuels the central high-$z$ galaxy via filaments, it is unclear how coherent these flows are upon entering the inner CGM. If the flows fragment, or form an extended structure around the galactic disc, it may result in delayed star formation, as opposed to hot-accreted gas which could reach the active star-forming regions more isotropically. To test this hypothesis, we study the components of the accretion geometry: the distance $r$ of a gas particle to the galactic centre, the height $h$ above the galactic plane\footnote{We define the galactic plane from the angular momentum of
stars orbiting the main galaxy at distances below $0.1\, R_ {\mathrm{vir}}$.}, the galactocentric/cylindrical radius $R_{\mathrm{c}}$, as well as the angle $\Theta$ between the galactic plane and the accreted material, subtended by $R_{\mathrm{c}}$ and $r$. Here an angle of \ang{0} corresponds to the gas orbiting the galaxy in the plane of the disc, while gas at (positive or negative) \ang{90} is accreted from above or below the central galaxy, perpendicular to the disc. Therefore, the angle distribution ranges from -\ang{90} to +\ang{90}, while for ideal isotropic accretion the median would be equal to \ang{0}. We present in \cref{fig:geometry} the median of $\Theta$ and of absolute values of $r$, $h$ and $R_{\mathrm{c}}$ up to \SI{600}{Myr} before star formation, for halo A. We also investigate the trajectories of 500 randomly selected tracer particles from each gas population, up to \SI{400}{Myr} before star formation, as shown in \cref{fig:fig_traj}. In \cref{pre}, we include trajectories of the three gas populations up to \SI{1.2}{Gyr} before star formation, in order to illustrate the structures from which the accreted gas originates: cold-accreted gas (both fast and slow) is accreted along filaments, while hot accretion is much more isotropic.

As seen in the left panel of \cref{fig:geometry}, up to \SI{600}{Myr} before star formation, slow gas resides within $0.25\, R_ {\mathrm{vir}}$, regardless if we look at $r$, $R_{\mathrm{c}}$ or $h$. On the other hand, the median distance $r$ at which fast gas resides is close to $2\, R_ {\mathrm{vir}}$ at \SI{600}{Myr} before star formation, before decreasing steadily until reaching the galaxy and arriving below $ 1/10\, R_ {\mathrm{vir}}$ only around \SI{100}{Myr} before star formation. The evolution of the angle $\Theta$ (right side of \cref{fig:geometry}) shows that {fast gas} arrives at the disc from a much steeper angle than slow gas, thus indicating that quick star formation can be achieved if the filaments provide gas from above or below the disc, rather than delivering it in the plane. Meanwhile, around \SI{600}{Myr} before star formation, hot-accreted gas is at around $1\, R_{\mathrm{vir}}$ away from the galaxy. From there, it reaches $0.1\, R_ {\mathrm{vir}}$ close to \SI{200}{Myr} before star formation, following a different trajectory than the `trapped' cold-accreted gas residing at $\sim 0.25\,R_{\mathrm{vir}}$.

Moreover, despite being at different distances away from the galaxy \num{200} to \SI{600}{Myr} before star formation, the densities of slow and fast gas are in a similar range, matching the values typical for streams of cold-accreted gas in the galactic halo (from \SI{0.03}{\mp.cm^{-3}} to \SI{0.6}{\mp.cm^{-3}}). Meanwhile, the hot-accreted gas arrives from a background with a much lower density, below \SI{0.03}{\mp.cm^{-3}}.

Combining these results with the trajectories in \cref{fig:fig_traj} reveals that slow gas is trapped in a turbulent accretion structure of density close to the one of cold gas streams, at distances $\sim 1/3-1/2\, R_ {\mathrm{vir}}$ (recall the travel distances presented in Figure~\ref{fig:fig_groups_dist}). There it lingers for a few hundred Myr before entering the dense disc formed below $ 1/10\, R_ {\mathrm{vir}}$, and later joining the region of active star formation. The fast cold gas penetrates the region below $ 1/3\, R_ {\mathrm{vir}}$ and quickly reaches the disc following straight trajectories, in contrast to the more complex flows in the slow gas. Meanwhile, hot-accreted gas, accreted from all directions quasi-spherically (as shown by the median $\Theta \sim \ang{10} - \ang{0}$ with 40 - 60th percentiles of around $\pm\ang{10}$ in agreement with an isotropic distribution \footnote{For comparison, the $40-\SI{60}{\percent}$ angle from purely isotropically distributed points on a sphere are $\pm \ang{11.6}$.}, \cref{fig:geometry,fig:fig_traj}), does not mix with the slow cold gas but rather arrives in the denser galactic disc below $ 1/10\, R_ {\mathrm{vir}}$ and forms stars in the central galaxy. In this picture, all three gas populations contribute to the angular momentum of the galaxy by depositing different amounts of it; the fast gas provides the fraction with the highest values by entering via angular-momentum-rich streams, while the angular momentum of cold-accreted gas in the turbulent region is lower due to higher levels of disordered motions and more efficient mixing of the gas, and so it arrives in the disc with smaller $j_{\mathrm{gas}}$, similarly to the isotropically accreted hot gas.

\section{Discussion and conclusions} \label{discus}

In this work, we use cosmological simulations of galaxy formation to investigate the behaviour of gas accretion and its relation to star formation in high redshift galaxies. This study is particularly timely as JWST is already providing observational data challenging what we currently know about star-forming galaxies at higher redshifts, thus increasing the need for more extensive studies of galaxy and star formation.

We find that cold-accreted gas --~defined here as gas that never heats up above \SI{3e4}{K} during accretion~-- travels faster than hot-accreted gas from the outer regions of dark matter haloes to the inner (\cref{fig:timescales_2_13}), in agreement with the literature (\eg{} \citealt{2005keres,2006MNRAS.368....2D, 2009Natur.457..451D, 2009MNRAS.397L..64A, 2012ApJ...745...11G}).
Although the total amount of mass that ends up forming stars remains dominated by cold streams above $z \sim 3.5-4$ (\cref{fig:fig_sf_rate}), this does not translate into cold streams fuelling star formation more quickly than hot gas accretion.
Indeed, by measuring the median timescale required for gas to form stars after entering the inner CGM ($1/3$ of the virial radius), $t_{1/3\shortarrow\star}$, we show that cold-accreted gas experiences a delay with respect to the hot-accreted one (\cref{fig:fig_combined}). The difference in median $t_{1/3\shortarrow\star}$ for these two populations of gas increases with time, reaching up to \SI{800}{Myr} at redshift $\sim 2-4$.

By splitting cold-accreted gas particles into two subgroups, slow and fast, based on whether their $t_{1/3\shortarrow\star}$ is smaller or larger than the median $t_{1/3\shortarrow\star}$ for the hot-accreted gas, we find that there are two ways in which cold-accreted gas can enter the galactic disc --- 1) a `fast' track, which feeds gas directly to the star-forming disc, and 2) a `slow' track, where gas enters a turbulent accretion region in the inner CGM that extends to approximately $0.5\,R_\mathrm{vir}$ (\cref{fig:fig_traj}). The majority of cold-accreted gas (\SI{90}{\percent} of the mass) is transported through this region, where no star formation occurs due to the low gas densities. This process takes several hundred million years, after which the gas reaches the central star-forming galaxy. In contrast, we find that hot-accreted gas typically fuels star formation much faster than cold accretion: it bypasses the turbulent region and is accreted directly into the star-forming disc from above and below the disc plane (\cref{fig:geometry}).

We conducted an analysis of possible explanations for this delay and found that longer $t_{1/3\shortarrow\star}$ correlates with shock-heating after accretion in the inner circumgalactic medium, as well as gas entrainment in galactic outflows. These factors contribute significantly to the delay, yet none of them explain it fully on their own. Removing particles which heat up above $T_{\mathrm{cold}} = \SI{3e4}{K}$ after crossing $ 1/3\, R_ {\mathrm{vir}}$ or which leave the inner CGM decreases the median value of $t_{1/3\shortarrow\star}$, but the median remains larger by a few hundred Myr than the median computed for hot-accreted gas (\cref{fig:fig_groups_temp,fig:fig_groups_dist}).

Furthermore, we find that slow gas takes longer to form stars despite having a lower $j_\mathrm{gas}$ than the fast gas in the few hundred Myr before star formation (\cref{fig:fig_AM}).
This allows us to rule out angular momentum as the driving factor behind the delay. While the cold-accreted gas coming from streams is rich in angular momentum, this is lost due to gas mixing upon entering the extended turbulent accretion region. Overall, the emerging picture is that the cold-accreted gas reaches the inner CGM quicker than hot-accreted gas. Here, however, a majority of the cold gas ends up trapped in an extended accretion region which causes a delay in star formation.

Our work shows that cold-accreted gas at $z > 2$ loses a significant amount of $j_{\mathrm{gas}}$ before star formation, in agreement with the findings of \cite{Danovich_2015}. In the inner CGM, cold-accreted gas is trapped for a few hundred Myr where both gravitational torques \citep{cadiou_GravitationalTorquesDominate_2021} and mixing with previously-accreted gas drive $j_\mathrm{gas}$ down. As a consequence, we expect longer delays to lead to a decorrelation of the spin of star-forming gas from its instantaneous value at accretion through the virial sphere, contributing to disconnecting the halo spin from the stellar one \citep{jiang_DarkmatterHaloSpin_2019,CC_stellar_angular}.
This work stresses the importance of studies based on the analysis of the Lagrangian history of the accreted gas to study the timescales involved in star formation and calls for further similar studies covering a statistical sample of galaxies spanning a larger redshift range.
This would notably allow us to establish the magnitude of this delay mechanism and its impact on the angular momentum properties of galaxies both at low redshifts \citep[][]{RomanowskyFall2012}, as well as at high redshifts relevant to the largest discs seen by JWST \citep[\eg{}][]{2023_kartaltepe}.

\section*{Acknowledgements}

We thank J.~Devriendt, A.~Slyz, A.~Dekel for their fruitful contributions, as well as C.~Pichon, Y.~Dubois, M.~Rey,  T.~Haworth and Á.~Segovia-Otero for providing feedback on the results obtained.

This work was supported by funding from the European Union's Horizon 2020 research and innovation programme under grant agreement No. 818085 GMGalaxies. The project used computing equipment funded by DiRAC (\SI{10}{Mcpu.hr}, project dp160). Data analysis was carried out on facilities supported by the Research Capital Investment Fund (RCIF) provided by UKRI and partially funded by the UCL Cosmoparticle Initiative. The work was also conducted using computing equipment available at the {Division of Astronomy, department of Physics} at Lund University. OA and CC acknowledge support from the Knut and Alice Wallenberg Foundation and the Swedish Research Council (grant 2019-04659). ZK is currently supported by Fulbright Graduate Award awarded by the Polish-U.S. Fulbright Commission (Scholarship Agreement No. PL/2023/2/GS).

The data analysis was conducted using \textsc{Python}, including \textsc{xarray} \citep{hoyer2017xarray}, \textsc{colossus} \citep{2018diemer}, \textsc{jupyter notebooks} \citep{2016jupyter}, \textsc{matplotlib} \citep{matplotlib_2007}, \textsc{numpy} \citep{Harris_2020}, \textsc{pynbody} \citep{pontzen_pynbody}, \textsc{tangos} \citep{Pontzen_2018_tangos,pontzen_tangos} and \textsc{yt} \citep{turk_2011}.

\section*{Author Contributions}

Using the CRediT (Contribution Roles Taxonomy) system (\url{https://authorservices.wiley.com/author-resources/Journal-Authors/open-access/credit.html}), the main roles of the authors were:\\
\noindent
\textbf{ZK}: conceptualisation; methodology; investigation; formal analysis; writing – original draft; visualisation.\\
\textbf{CC}: conceptualisation; methodology; supervision; data curation; resources; writing – review and editing; validation.\\
\textbf{OA}: conceptualisation; writing – review and editing; resources; funding acquisition; validation.\\
\textbf{AP}: conceptualisation; writing – review and editing; resources; validation.

%%%%%%%%%%%%%%%%%%%%%%%%%%%%%%%%%%%%%%%%%%%%%%%%%%
\section*{Data Availability}

The data underlying this article will be shared on reasonable request to the corresponding author.



\bibliographystyle{mnras}
\bibliography{authors} % if your bibtex file is called example.bib

%\begin{thebibliography}{99}
%\end{thebibliography}

%%%%%%%%%%%%%%%%%%%%%%%%%%%%%%%%%%%%%%%%%%%%%%%%%%

%%%%%%%%%%%%%%%%% APPENDICES %%%%%%%%%%%%%%%%%%%%%

\appendix

\section{The `NUT' simulation} \label{NUT}

\begin{figure}
	\includegraphics[width=\columnwidth]{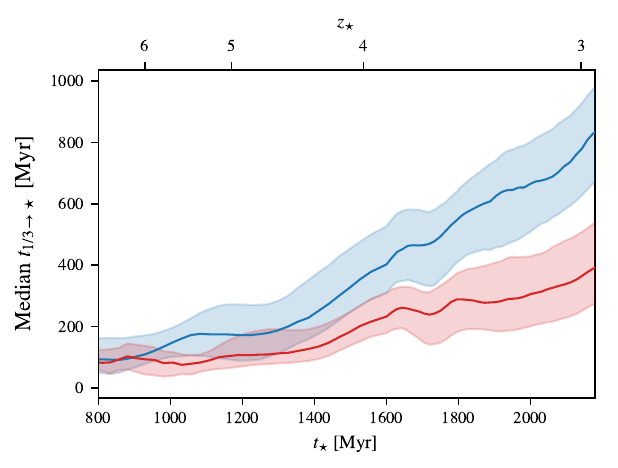}
    \caption{Same as \cref{fig:fig_combined} but for the NUT simulation. The median conversion timescale $t_{1/3\shortarrow\star}$ cold-accreted gas required to form stars after crossing $ 1/3\, R_ {\mathrm{vir}}$ for the first time is higher than the median for hot-accreted gas; same as in the New Horizon simulations, cold-accreted gas experiences a delay in forming new stars.}
    \label{fig:NUT_13_SF}
\end{figure}

Using NUT, we conduct our analysis over the redshift range $ 3 < z < 8 $, where the halo mass at redshift $z \sim 3$ is \SI{8.5e10}{\Msun}. Since the halo mass is lower in NUT than in the rest of the simulations studied, we expect the contribution to accretion from the hot mode to be lower at $z \sim 2$ than in Halo A, B and C.

The simulation reaches a maximum physical spatial resolution of $\approx \SI{10}{pc}$ at all times, with comoving box size of \SI{9}{Mpc.\hh^{-1}}. Additional technical details about this simulation, which include a description of the star formation recipe implemented, are provided in the Appendix in \citet{powell_ImpactSupernovadrivenWinds_2011} and \citet{Rams_y_2021}.

In \cref{fig:NUT_13_SF}, we show the median conversion timescale $t_{1/3\shortarrow\star}$ for hot- and cold-accreted gas in this simulation. These two subsets comprise \SI{60}{\percent} of the total accreted gas mass over time, at $2 < z < 15$. We conducted the same analysis to investigate the source of the delay in this simulation as we did using the New Horizon simulations, which provided the same conclusions --- the majority of cold-accreted gas settles in an extended structure of accretion before joining the active star-forming region.
However, as mentioned above, the NUT simulation has no SN feedback included so the delay visible here cannot be caused by particles ejected from the inner CGM by supernova explosions (\SI{85}{\percent} of cold-accreted gas stays inside the inner CGM after entering it). Moreover, the galaxy resides in a more isolated environment with no major mergers happening, hence the lines grow steadily without major disruptions with star formation time. The maximum difference between the two medians is close to \SI{500}{Myr} and is reached at redshift $\sim 3$.

\section{Pre-accretion trajectory of gas} \label{pre}

\begin{figure*}
	\includegraphics[width=\linewidth]{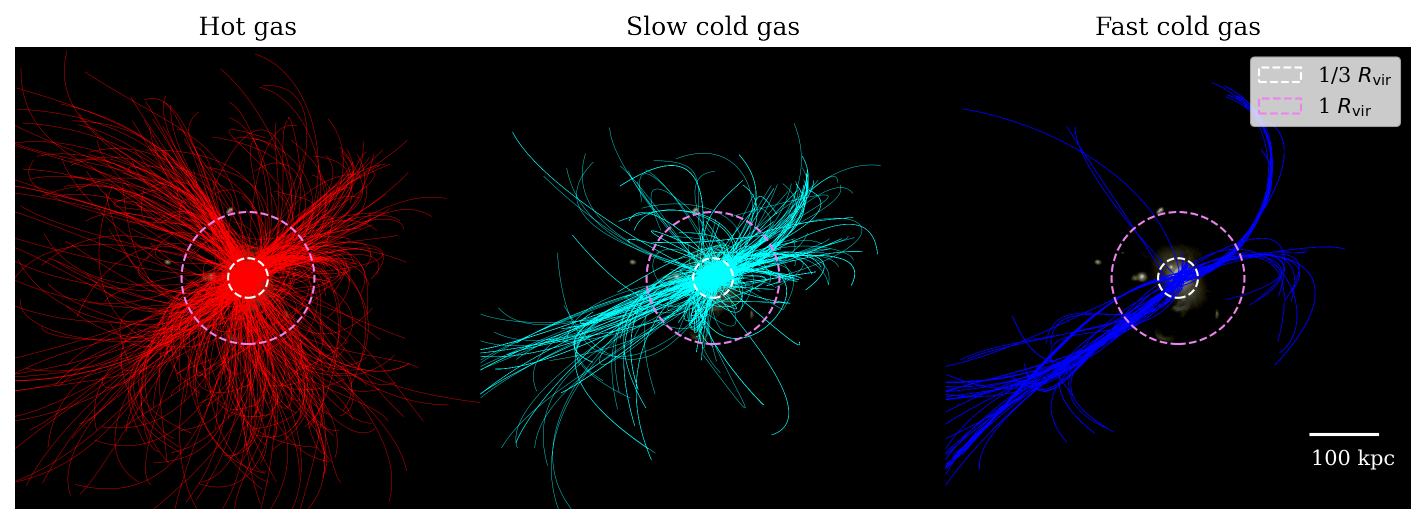}
    \caption{Same as in \cref{fig:fig_traj}, except here we show tracer positions up to \SI{1.2}{Gyr} before star formation. Hot-accreted gas (left) enters the inner CGM nearly isotropically, albeit with some alignemnt with large-scale filaments, while both slow-cold gas (middle) and fast-cold gas (right) originate from the dense matter filaments and travel through the galactic halo along the cold streams. All the tracer particles presented have $t_{\mathrm{\star}}$ between $2 < z < 3$. Dashed circles show the radii $ 1/3\, R_ {\mathrm{vir}}$ and $ R_{\mathrm{vir}}$.}
    \label{fig:fig_traj_pre}
\end{figure*}
We investigate the origin of the slow cold, fast cold and hot-accreted gas by plotting the trajectories of 500 randomly selected tracer particles from each subset up to \SI{1.2}{Gyr} before star formation, as seen in \cref{fig:fig_traj_pre}. The trajectories confirm that both slow and fast gas accrete in a filamentary fashion, whereas hot gas is accreted quasi-spherically.

%%%%%%%%%%%%%%%%%%%%%%%%%%%%%%%%%%%%%%%%%%%%%%%%%%

% Don't change these lines
\bsp	% typesetting comment
\label{lastpage}
\end{document}